%% file: main.tex
\documentclass[conference, compsoc]{IEEEtran}
\PassOptionsToPackage{hyphens}{url}
\usepackage{hyperref}
\usepackage{algorithm}
\usepackage{algpseudocode}
\usepackage{dirtytalk}
\usepackage{graphicx}
\usepackage{amsmath}

\def\BibTeX{{\rm B\kern-.05em{\sc i\kern-.025em b}\kern-.08em
    T\kern-.1667em\lower.7ex\hbox{E}\kern-.125emX}}

\begin{document}

\title{Reinforcing Security and Usability of Crypto-Wallet with Post-Quantum Cryptography and Zero-Knowledge Proof}

\newcommand{\old}[1]{}
\newcommand{\quotes}[1]{``#1''}

\author{
\IEEEauthorblockN{
Yathin Kethepalli \IEEEauthorrefmark{1}, 
Rony Joseph \IEEEauthorrefmark{1}, 
Sai Raja Vajrala \IEEEauthorrefmark{1}, 
Jashwanth Vemula \IEEEauthorrefmark{1}, 
and 
Nenavath Srinivas Naik \IEEEauthorrefmark{2}
}

\IEEEauthorblockA{
\IEEEauthorrefmark{1}IIIT Naya Raipur,
\IEEEauthorrefmark{2}IIITDM Kurnool
}
}

\IEEEoverridecommandlockouts

\maketitle

\begin{abstract}
    Crypto-wallets or digital asset wallets are a crucial aspect of managing cryptocurrencies and other digital assets such as NFTs. However, these wallets are not immune to security threats, particularly from the growing risk of quantum computing. The use of traditional public-key cryptography systems in digital asset wallets makes them vulnerable to attacks from quantum computers, which may increase in the future. Moreover, current digital wallets require users to keep track of seed-phrases, which can be challenging and lead to additional security risks.
    To overcome these challenges, a new algorithm is proposed that uses post-quantum cryptography (PQC) and zero-knowledge proof (ZKP) to enhance the security of digital asset wallets. The research focuses on the use of the Lattice-based Threshold Secret Sharing Scheme (LTSSS), Kyber Algorithm for key generation and ZKP for wallet unlocking, providing a more secure and user-friendly alternative to seed-phrase, brain and multi-sig protocol wallets. This algorithm also includes several innovative security features such as recovery of wallets in case of downtime of the server, and the ability to rekey the private key associated with a specific username-password combination, offering improved security and usability. The incorporation of PQC and ZKP provides a robust and comprehensive framework for securing digital assets in the present and future. This research aims to address the security challenges faced by digital asset wallets and proposes practical solutions to ensure their safety in the era of quantum computing.
\end{abstract}


\section{Introduction}
    
    Digital asset wallets have emerged as an indispensable tool for the managing cryptocurrencies. These are designed to store and manage various digital assets, including but not limited to Bitcoin, Ethereum, and other cryptocurrencies. Digital asset wallets have revolutionized the traditional concept of wallets by providing a secure and convenient way to manage digital assets, thus enabling users to engage in various financial activities. The origin of digital asset wallets can be traced back to the introduction of Bitcoin, the first-ever cryptocurrency that was introduced in 2009 \cite{satoshi2009bitcoin}. The advent of Bitcoin led to the introduction of the blockchain technology, which allowed for the storage of transactions on a public ledger. These wallets are designed to store private keys required for accessing the blockchain and conducting transactions. In today's digital age, they serve a variety of purposes, including but not limited to, sending and receiving cryptocurrency, managing portfolios, investing in blockchain projects, and as a means of payment. They have gained immense popularity due to their security features, which are far more advanced than traditional wallets. Most of these wallets incorporate multi-signature authentication, and cold storage mechanisms to protect the user's funds from any malicious attacks.

    Various types of digital asset wallets are available today, each with its own set of features and security protocols. These include software, hardware, paper, brain and smart contract wallets \cite{erinle2023sok}.
    \begin{itemize}
        \item \textbf{Software Wallets:} These wallets are installed on a user's computer or mobile device or accessed through a web browser. They are generally easy to use, but are more susceptible to online hacks and attacks.
        \item \textbf{Hardware Wallets:} These are physical devices that are used to store digital assets offline. They are generally more secure than software wallets, but can be more difficult to use.
        \item \textbf{Paper Wallets:} These are physical pieces of paper that contain a user's private key. They are often used as a backup for software or hardware wallets.
        \item \textbf{Brain Wallets:} Wallets created by deriving a private key from a passphrase chosen by the user. They can be accessed online from any device but are less secure due to vulnerability to brute-force and dictionary attacks.
        \item \textbf{Smart contract Wallets:} These wallets utilize smart contracts to securely manage assets and provide advanced customization features. These wallets offer enhanced security and enable functionalities like recoverable wallets, signless transactions, and batched transactions, surpassing the capabilities of traditional crypto wallets.
    \end{itemize}

    Digital asset wallets have emerged as a critical component in the rapidly evolving world of digital finance. Cryptography has become an essential component of these wallets, providing a secure means of storing and transferring digital assets. Mathematical algorithms form the backbone of modern cryptography, ensuring that these wallets remain secure and accessible only to those with the appropriate keys. Cryptography plays a crucial role in securing digital asset wallets by converting sensitive information into an unreadable format that can only be deciphered using the correct decryption keys. The importance of cryptography in digital asset wallets cannot be overstated, as it protects the user's funds from theft and malicious attacks. The use of cryptographic techniques, such as public-key cryptography and hash functions ensures that digital assets are stored and transferred securely, without the need for a central authority. The user's private keys are used to ensure that only authorized individuals have access to their digital assets.

    The advent of quantum computing presents a formidable obstacle to modern cryptographic systems, which include digital asset wallets. While classical computers rely on binary digits or bits to represent information, quantum computers leverage the principles of quantum mechanics to use quantum bits or qubits. These qubits can exist simultaneously in multiple states, allowing quantum computers to perform calculations in parallel and potentially lead to exponential speedups for specific problems. This new technology has the potential to revolutionize multiple fields, such as cryptography and security. However, one of the primary implications of quantum computing for security lies in the breaking of cryptographic systems that rely on the complexity of specific mathematical problems, such as factoring large numbers or computing discrete logarithms. Most widely used cryptographic protocols, such as RSA and ECC, are based on these problems' difficulty and can be compromised by Shor's algorithm on a sufficiently powerful quantum computer \cite{shor1997polynomial, bernstein2009quantum}. Consequently, it is widely believed that numerous cryptographic systems currently in use will become vulnerable once large-scale quantum computers are available. Thus, the rise of quantum computing has introduced significant security concerns that require immediate attention. As researchers continue to explore new ways to counter the threat, it is essential to acknowledge the potential limitations of current cryptographic systems and plan for more secure alternatives.
    
    Post-Quantum Cryptography (PQC) focuses on the design and analysis of cryptographic protocols that are resistant to attacks by classical and quantum computers. The primary objective is to devise cryptographic systems that can withstand attacks by quantum computers, which are expected to become increasingly powerful in the future. As the development of PQC is still in its nascent stages, several cryptographic schemes are presently being evaluated for their effectiveness against quantum attacks.

    One of the cryptographic concepts that have gained considerable attention in recent times is zero-knowledge proof (ZKP). It is a technique that enables one party to prove to another that a given statement is true, without revealing any additional information beyond the veracity of the statement. In other words, ZKP enables authentication without the need to disclose any private information.

    Our proposed methodology aims to address the security concerns associated with traditional public-key cryptography systems and seed-phrases in digital asset wallets by employing PQC and ZKP. It includes several innovative security features that enhance the security of digital asset wallets and makes them easier to use. One such feature is recovery of wallets in case of server downtime, which allows users to recover access to their wallets even if the server isn't accessible for any reason. This approach is more secure and less cumbersome than traditional recovery mechanisms that rely on email or phone verification. Another feature is the ability to rekey the private key associated with a specific username-password combination, which provides an additional layer of security in case a user's private key is compromised.

    The rest of this paper is organized as follows: Section \ref{sec:related-works} presents related works on crypto wallets, Section \ref{sec:background} provides the necessary background information for understanding our proposed methodology. In Section \ref{sec:proposed-methodology}, we provide a detailed explanation of our proposed methodology, which employs PQC and ZKP to enhance the security of digital asset wallets. We also discuss the innovative security features of our approach, including client-side private key derivation, social recovery, and rekeying. Section \ref{sec:results} presents the results of our experimental testing, which demonstrates the effectiveness of our approach in improving digital asset wallet security. We consider the potential impact of our approach on the adoption of digital wallets and the broader digital finance industry. We also discuss the limitations of our approach and the challenges that remain to be addressed. Section \ref{sec:implications} discusses the broader implications of our work for the future of digital asset wallet security. Finally, in Section \ref{sec:conclusion}, we conclude the paper and discuss directions for future research.

\section{Related Works}
\label{sec:related-works}
Over the years, various research studies have been conducted to investigate diverse aspects of digital asset wallets. These studies have examined both general and specific aspects of the subject, highlighting the multidimensional nature of research in this domain. The focus of these studies can be broadly classified into three categories: key management, wallet design, and wallet security.

\subsection{Key Management}
The security of digital asset wallets relies heavily on robust key management practices, which have been extensively studied in the literature \cite{mangipudi2022uncovering, he2018social, pal2021key, he2019novel, courtois2017stealth, Belchior2022, Chen2020b, suratkar2020cryptocurrency}.

Key management encompasses a range of methods for securely handling and storing private keys, which can be implemented across various digital and physical applications and distributed among multiple devices. Mangipudi et al. \cite{mangipudi2022uncovering} categorized wallets into single-device and multi-device wallets based on their ability to distribute the risk associated with private key compromise. Expanding on this notion, Pal et al. \cite{pal2021key} explored additional avenues for key management, such as cloud-based key storage.

Furthermore, researchers have delved into key management approaches specific to individual assets \cite{eskandari2018first}, specialized hardware devices like Hardware Security Modules (HSMs) \cite{gotte2021tech, shbair2021hsm}, and innovative key recovery methods \cite{he2019novel}. These studies underscore the significance of effective key management in crypto-wallets and highlight diverse strategies for safeguarding private keys.

\subsection{Wallet Design}
The design of cryptocurrency wallets has garnered significant attention in research, particularly in terms of wallet architecture. Khadzhi et al. \cite{khadzhi2020method} conducted an analysis of transaction activity on hardware wallets, shedding light on their usage patterns. Khan et al. \cite{khan2019security} proposed a novel architectural design that leverages QR codes to authenticate transactions between hardware and software wallets. Eyal \cite{eyal2022cryptocurrency} offered valuable insights into wallet design by assessing the failure probability of wallets from a security standpoint. Furthermore, specific studies have explored the design characteristics of Ethereum wallets \cite{di2020characteristics, homoliak2018air}.

A notable architectural concept highlighted in the literature is the main and sub-wallet setup, akin to a savings and current account structure for blockchain wallets. Rezaeighaleh and Zou \cite{rezaeighaleh2019deterministic} discussed the concept of sub-wallets and main-wallets within deterministic wallets and developed a prototype implementation in a hardware wallet. This mechanism was found to be resistant to classical wallet attacks, including Man-In-The-Middle attacks, where malicious actors attempt to substitute a user's address with a malicious one.

These studies underscore the importance of thoughtful architectural considerations in cryptocurrency wallet design, paving the way for innovative solutions that enhance security and user experience.

\subsection{Wallet Security}
The security and vulnerability of blockchain systems have been extensively examined in the literature \cite{chen2020survey, li2020survey, guo2022survey, zamani2020security, Urien2021}. Notably, Homoliak et al. \cite{homoliak2018air} conducted a comprehensive analysis of attacks on crypto wallet applications and proposed defense mechanisms to mitigate vulnerabilities. Their study emphasizes the adoption of air-gapped hardware wallet solutions as an effective defense measure.

Addressing server-side vulnerabilities, Bui et al. \cite{bui2019pitfalls} specifically focus on wallet attack scenarios and propose defense mechanisms, particularly against man-in-the-middle attacks. Another study explores prevention mechanisms for phishing attacks, a prevalent type of wallet attack \cite{andryukhin2019phishing}. Meanwhile, Chen et al. \cite{chen2020survey} delve into the vulnerabilities, attacks, and security mechanisms specific to the Ethereum blockchain. Their research contributes to the existing body of knowledge by shedding light on vulnerabilities and attacks across different layers of the blockchain, with a particular focus on wallet-related threats in the application layer.

These studies collectively contribute to enhancing wallet security in blockchain systems by identifying vulnerabilities, proposing defense mechanisms, and highlighting the importance of robust security practices throughout the ecosystem.

\section{Technical Background}
\label{sec:background}

    \subsection{Post-Quantum Cryptography}
    \old{https://en.wikipedia.org/wiki/Post-quantum_cryptography#Implementation}
    PQC pertains to cryptographic techniques, primarily public-key algorithms, which are conjectured to resist cryptanalytic attacks by quantum computers. The imminent threat posed by quantum computing towards traditional cryptographic protocols, which depend on the computational complexity of certain mathematical problems, most notably, integer factorization, the discrete logarithm, and the elliptic-curve discrete logarithm problems, have engendered the necessity for the development of PQC.

    The significance of PQC in ensuring secure communication is paramount in light of recent advancements in quantum computing technology. The development and deployment of PQC systems and protocols, which are both robust and efficient, is an active area of research in the field of cryptography. It is imperative to ensure that the cryptographic protocols of the future are resilient against the potent attacks of quantum computers, and PQC stands at the forefront of this endeavor.

    \input{Tables/PQC-Comparision}

    A ubiquitous trait of numerous PQC algorithms is their require larger key sizes than the conventional pre-quantum public key algorithms. Such algorithms present a conundrum in key size, computational efficiency, and ciphertext or signature size, warranting careful consideration of the trade-offs between these factors. The Table \ref{tab:PQC-Comparision} presents key sizes, ciphertext and signature sizes for various schemes operating at a post-quantum security level of 128 bits, thereby enabling an assessment of these trade-offs. Currently, PQC research is mostly focused on six different approaches using lattices, multivariate equations, hashes, codes, supersingular elliptic curve isogeny and symmetric keys \cite{gershon2013qubit, spectrum2008quantum}.

        \subsubsection{Lattice-based cryptography} This approach involves the use of systems such as learning with errors, ring learning with errors (ring-LWE) \cite{bernstein2009sharcs, bernstein2010gvm, cryptoeprint:2014/070}, ring learning with errors key exchange, and ring learning with errors signature. It also includes the older NTRU or GGH encryption schemes and the newer NTRU signature and BLISS signatures \cite{10.1007/978-3-642-33027-8_31}. Notably, some schemes like NTRU encryption have been subjected to extensive studies without yielding any feasible attack. In contrast, others like the ring-LWE algorithms have security proofs that reduce to a worst-case problem \cite{10.1007/978-3-662-46803-6_24}. The PQC Study Group sponsored by the European Commission has recommended the Stehle–Steinfeld variant of NTRU for standardization instead of the NTRU algorithm because it has a security reduction, whereas the latter is still patented \cite{cryptoeprint:2013/383, 10.1145/2535925}. Studies have indicated NTRU may have more secure properties than other lattice-based algorithms \cite{Augot2015InitialRO}. Ring-LWE, a specific version of Ring-LWE signatures, has security reduction to the shortest-vector problem (SVP) in a lattice known to be NP-hard. Furthermore, the security of the NTRU encryption scheme and the BLISS signature is believed to be related to the Closest Vector Problem (CVP) in a Lattice, also known as NP-hard. However, it cannot be provably reducible.

        \old{
        \subsubsection{Multivariate cryptography} pertains to using cryptographic systems grounded in the Rainbow (Unbalanced Oil and Vinegar) scheme, which hinges on the complexity of solving multivariate equation systems. Although numerous efforts to establish multivariate equation encryption schemes have been unsuccessful, multivariate signature schemes such as Rainbow hold the potential to serve as the basis for a digital signature that can withstand the onslaught of quantum computing \cite{cryptoeprint:2013/004}. It is worth noting that the Rainbow Signature Scheme is subject to a patent. Unbalanced Oil and Vinegar signature schemes are classified as asymmetric cryptographic primitives based on multivariate polynomials over a finite field. Bulygin, Petzoldt, and Buchmann have demonstrated that generic multivariate quadratic UOV systems can be reduced to the NP-Hard Multivariate Quadratic Equation Solving problem, thereby contributing to the current understanding of the limitations and possibilities of this approach \cite{cryptoeprint:2016/030}.

        \subsubsection{Hash-based cryptography} It is a type of digital signature scheme that uses hash functions to provide security. One of the most well-known hash-based signature schemes is the Merkle signature scheme, which was invented by Ralph Merkle in the late 1970s. Although interest in these signatures waned due to their limited capacity, they have gained renewed interest due to the potential for quantum computers to break traditional number-theoretic digital signatures like RSA and DSA. The Merkle signature scheme has a security reduction to the underlying hash function \cite{10.1145/3292548, 10.1007/978-3-642-17401-8_3}. It has been recommended by the PQC Study Group sponsored by the European Commission for long term security protection against quantum computers \cite{cryptoeprint:2013/383}. Other hash-based signature schemes include Lamport signatures, XMSS \cite{8666459, 10.1007/978-3-642-25405-5_8}, SPHINCS \cite{10.1007/11496137_12} and WOTS schemes.

        \subsubsection{Code-based cryptography} This subfield leverages error correcting codes as the foundation of cryptographic schemes. Notable code-based cryptosystems include the McEliece and Niederreiter encryption algorithms and the Courtois, Finiasz, and Sendrier signature scheme. The original McEliece signature has demonstrated resilience over 40 years, despite many attempts to introduce structural modifications to the underlying code to improve key size \cite{rfc8391}. However, some of these variations are insecure. The McEliece encryption system has been recommended by the PQC Study Group sponsored by the European Commission as a viable candidate for long-term protection against quantum computer attacks. It has a security reduction to the Syndrome Decoding Problem, known as NP-hard \cite{cryptoeprint:2013/383, PEREIRA201695}.

        \subsubsection{Supersingular elliptic curve isogeny cryptography} It is an innovative cryptographic paradigm that leverages the properties of supersingular elliptic curves and supersingular isogeny graphs to create a promising alternative to the prevailing Diffie–Hellman and elliptic curve Diffie–Hellman key exchange methods, which are susceptible to quantum computing attacks \cite{10.1145/73007.73011}. The system's ability to provide forward secrecy, a key attribute for preventing government mass surveillance and safeguarding against key failure-induced long-term key compromise, underscores its practical utility \cite{Overbeck2009}. The system's security derives from the challenge of constructing an isogeny between two supersingular curves with the same number of points, which is considered a hard computational problem. The recent work by Delfs and Galbraith further bolsters the case for the cryptographic system's robustness, as the problem was as arduous as suggested by its inventors, and no security reduction to any known NP-hard problem exists. The cryptographic system's potential is not limited to key exchange, as Sun, Tian, and Wang demonstrated its suitability for creating quantum-resistant digital signatures based on supersingular elliptic curve isogenies \cite{10.1007/978-3-642-25405-5_2}. Furthermore, the cryptographic system's intellectual property is in the public domain, as no patents cover it.

        \subsubsection{Symmetric key quantum resistance} AES and SNOW 3G are already resistant to attack by a quantum computer, as long as sufficiently large key sizes are used \cite{higgins2013webprivacy}. Furthermore, key management systems and protocols that use symmetric key cryptography, such as Kerberos and the 3GPP Mobile Network Authentication Structure, are inherently secure against attack by a quantum computer. Some researchers suggest expanding symmetric key management, such as Kerberos, as an efficient way to implement PQC today, given its widespread worldwide deployment \cite{6337933}.}

    \subsection{Zero-Knowledge Proofs}
    \old{https://en.wikipedia.org/wiki/Zero-knowledge_proof#Blockchains}
    ZKPs are cryptographic protocols that enable one party (the prover) to convince another party (the verifier) that a statement is true, without revealing any additional information about the statement or any secrets used to prove it. The fascinating aspect of ZKPs is that it is trivial to demonstrate possession of certain knowledge by merely revealing it \cite{lixie2017zeroknowledge}. However, the challenge lies in proving possession of such knowledge without disclosing the knowledge itself or any other related information. In this sense, they aim to achieve high privacy and confidentiality by not exposing sensitive data beyond the necessary proof of knowledge. These proofs can be interactive, where both parties engage in a back-and-forth conversation, or non-interactive, where a single message generates the proof, commonly relying on the Fiat-Shamir heuristic \cite{10.1145/62212.62222}. Nonetheless, the validity of the proof still relies on computational assumptions and cryptographic primitives \cite{Wu2014, 10.1007/0-387-34805-0_60}. ZKPs have gained tremendous attention due to their potential applications in many fields, including cryptocurrency, blockchain, and cybersecurity.

    The three main properties of a ZKP are as follows:
    \begin{itemize}
        \item \textbf{Completeness }requires that if the statement is true, then a verifier will be convinced of its truth by an honest prover.
        \item \textbf{Soundness }requires that if the statement is false, no cheating prover can convince an honest verifier that it is true, except with some small probability.
        \item \textbf{Zero-knowledge }requires that if the statement is true, no verifier can learn anything other than the fact that the statement is true. This is formalized by showing that every verifier has some simulator that can produce a transcript that \quotes{looks like} an interaction between an honest prover and the verifier in question.
    \end{itemize}
    The initial two properties are applicable to a broader class of interactive proof systems. However, it is the third property that uniquely characterizes a ZKP. This distinctive quality ensures that the verifier gains no additional knowledge beyond the fact that the statement is true, even though the prover provides convincing evidence of their knowledge of the secret. This feature sets ZKPs apart from other interactive proof systems \cite{10.1007/3-540-39118-5_13}.

    Different variants of ZKPs can be defined based on how closely the distributions produced by the simulator and the real proof protocol match. Perfect ZKPs require that the distributions are the same, while statistical ZKPs allow for some statistical difference between the two distributions. Computational ZKPs go even further by requiring no efficient algorithm to distinguish between the two distributions. These variants are important because they allow for different levels of security and efficiency depending on the specific use case.

    \subsection{Kyber Algorithm}
    \old{Draw tables later: https://pq-crystals.org/kyber/}
    
    \begin{figure}[h]
		\centering\includegraphics[scale=0.8]{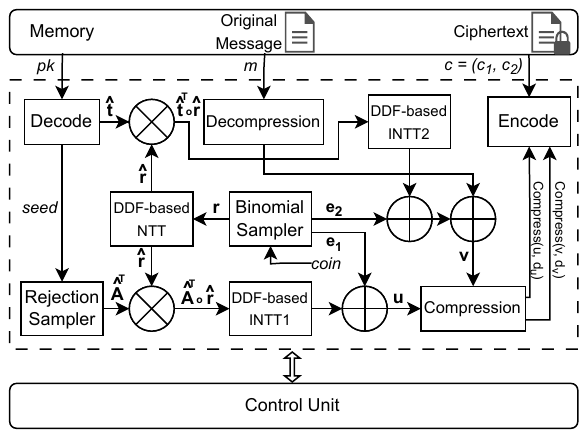}
		\caption{Working mechanism of Kyber algorithm}
     \label{fig:kyber}
    \end{figure}

    The Kyber algorithm is a sophisticated lattice-based cryptographic scheme that leverages the learning with errors (LWE) problem to establish a secure communication channel between multiple parties. The LWE problem entails detecting a concealed vector within a noisy vector, which is a challenging mathematical problem, and uses module LWE with cyclotomic rings as its trapdoor function. The Kyber algorithm generates both a public key and a private key, which are instrumental in facilitating the encryption and decryption processes, respectively, as shown in Figure \ref{fig:kyber}. The construction of these keys is grounded on matrices and polynomials derived from lattices and geometric structures utilized to create complex mathematical problems.

    The Kyber algorithm crafts the public key by generating a secret vector with coefficients from a finite field. The next step is to develop a matrix with random coefficients and multiply it by the secret vector. The resulting matrix undergoes rounding to the closest multiple of a certain number, ultimately constituting the public key. The public key is essentially the fundamental matrix constructed from the lattice basis. To form the private key, the Kyber algorithm crafts a matrix that comprises random coefficients and multiplies it by the public key. The resulting matrix is rounded to the nearest multiple of a certain number, which produces the private key. Like the public key, the private key is a fundamental matrix constructed from the same lattice basis as the public key matrix.

    The Kyber family of cryptographic algorithms delivers varying degrees of security, with Kyber-512 presenting a level comparable to AES-128, Kyber-768 having a security level equivalent to AES-192, and Kyber-1024 providing a security level that is comparable to AES-256. The security of the Kyber algorithm rests on the hardness of the LWE problem and the utilization of lattices in constructing the keys and ciphertext.
    
    \subsection{SHA256}
    Secure Hash Algorithms (SHA) are a family of cryptographic hash functions created by the United States National Security Agency (NSA). While there are other variants such as SHA 224 and SHA 384, SHA 256 has been at the forefront of real-world applications. SHA256 function produces a 256-bit value, typically a 64-digit hexadecimal number using 64 rounds of mixing and compression as shown in Algorithm \ref{algo:sha256}. It takes an input message of any length and produces a fixed-length output hash.

    SHA 256 is widely utilized in various applications where data integrity and security are paramount. Its significance extends to digital signature algorithms, message authentication codes, and notably, blockchain networks such as Bitcoin, Ethereum. It plays a fundamental role in ensuring immutability and tamper resistance of data within blockchain systems. It accomplishes this by generating unique hash values for each block in the blockchain, securing transactions and maintaining the integrity of the entire chain.

    \begin{algorithm}
	\caption{SHA256 function}
	\label{algo:sha256}
	\begin{algorithmic}
		\Function{SHA256}{$message$}
		\State $h_0 \gets \textit{0x6a09e667}$, $h_1 \gets \textit{0xbb67ae85}$, $h_2 \gets \textit{0x3c6ef372}$, 
		\State $h_3 \gets \textit{0xa54ff53a}$, $h_4 \gets \textit{0x510e527f}$, $h_5 \gets \textit{0x9b05688c}$, 
        \State $h_6 \gets \textit{0x1f83d9ab}$, $h_7 \gets \textit{0x5be0cd19}$
		
		\State $k_i \gets \lfloor 2^{32} \times |\sin i|\rfloor$ \texttt{for} $i$ \texttt{in} $0$ \texttt{to} $63$
		
		\State $a \gets h_0$, $b \gets h_1$, $c \gets h_2$, $d \gets h_3$,  
        \State $e \gets h_4$, $f \gets h_5$, $g \gets h_6$, $h \gets h_7$
		
		\For{$i$ in $1$ to $N$}
		\If{$0 \leq j < 16$}
		\State $M_{i,j}$
		\Else
		\State $\sigma_1(w_{j-2}) + w_{j-7} + \sigma_0(w_{j-15}) + w_{j-16}$
		\EndIf
		\For{$j$ in $0$ to $63$}
		\State $T_1 \gets h + \Sigma_1(e) + \mathcal{Ch}(e, f, g) + k_j + w_j$
		\State $T_2 \gets \Sigma_0(a) + \mathcal{Maj}(a, b, c)$
		\State $h \gets g$, $g \gets f$, $f \gets e$,
		\State $e \gets d + T_1$,
		\State $d \gets c$, $c \gets b$, $b \gets a$,
		\State $a \gets T_1 + T_2$
		\EndFor
		\State $h_0 \gets h_0 + a$, $h_1 \gets h_1 + b$, $h_2 \gets h_2 + c$, $h_3 \gets h_3 + d$,
		\State $h_4 \gets h_4 + e$, $h_5 \gets h_5 + f$, $h_6 \gets h_6 + g$, $h_7 \gets h_7 + h$
		\EndFor
		
		\State $hash \gets$ concatenate $h_0$, $h_1$, $h_2$, $h_3$, $h_4$, $h_5$, $h_6$, $h_7$
		
		\State \Return $hash$
		\EndFunction
	\end{algorithmic}
    \end{algorithm}
    
    The working of SHA256 involves the following steps:
    \begin{itemize}
        \item \textbf{Padding:} The input message is padded to make its length a multiple of 512 bits. Padding is added so the resulting padded message length is congruent to 448 modulo 512. The last 64 bits are used to store the original message length.
        \item \textbf{Message schedule:} The padded message is divided into 512-bit blocks, and a message schedule is created for each block. The message schedule consists of 64 32-bit words.
        \item \textbf{Initial hash values:} The initial hash values are pre-defined for each round of the SHA256 algorithm. There are eight hash values, each represented by a 32-bit word.
        \item \textbf{Compression function:} The compression function is applied to each message schedule and the initial hash values to produce a new set of hash values. The compression function involves several rounds of logical operations and also uses a set of pre-defined constant values.
        \item \textbf{Final hash:} After all the blocks have been processed, the final hash value is produced by concatenating the eight 32-bit hash values produced in the compression function.
    \end{itemize}

    \subsection{BCrypt}
    \old{https://en.wikipedia.org/wiki/Bcrypt}

    BCrypt is a widely used password-hashing function based on the Blowfish cipher. It provides a strong level of security and is known for its adaptability and resistance to brute-force attacks. It is designed to be computationally expensive, making it time-consuming and resource-intensive for attackers attempting to crack hashed passwords. One of the key features of BCrypt is its adaptive nature. It adjusts the computation complexity based on a configurable iteration count, known as the cost parameter. By increasing the cost, the hash function requires more time and computational resources to compute the hash value, effectively slowing down potential attackers. This adaptive nature allows BCrypt to remain resilient against evolving hardware capabilities and increases the security of hashed passwords over time.
    To further enhance security, it incorporates a salt value into the hashing process. The salt is a randomly generated string that is combined with the password before hashing. This ensures that even if two users have the same password, their hash values will be different due to the unique salt values. The inclusion of salt protects against rainbow table attacks, where an attacker precomputes the hash values of commonly used passwords to quickly identify matches.

    \begin{algorithm}[H]
        \caption{BCrypt function}
        \label{algo:bcrypt}
        \begin{algorithmic}
        \Function{bcrypt}{$cost, salt, message$}
            \State $P, S \gets \Call{EksBlowfishSetup}{cost, salt, message}$
            \State $ctext \gets$ \quotes{Any Input Message}
            \For{$i \gets 0$ to $63$}
                \State $ctext \gets \Call{EncryptECB}{P, S, ctext}$
            \EndFor
            \State $hash \gets \Call {Concatenate}{cost, salt, ctext}$
            \State \Return $hash$
        \EndFunction
        \end{algorithmic}
    \end{algorithm}
    
\begin{figure}[!h]
		\centering\includegraphics[scale=0.65]{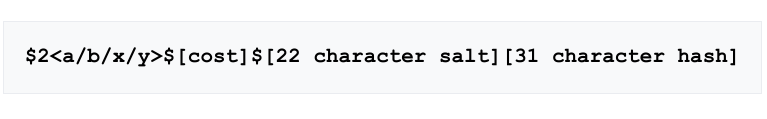}
		\caption{Final output of BCrypt function}
     \label{fig:bcrypt}
    \end{figure}

    It has gained significant popularity and has been implemented in various programming languages and frameworks, including C, C++, Java, JavaScript, Python, and Ruby. It is the default password hash algorithm for OpenBSD, a highly secure operating system.
    The input to the bcrypt function includes the message string (up to 72 bytes), a numeric cost parameter, and a 16-byte salt value as shown in Algorithm \ref{algo:bcrypt}. The cost parameter determines the number of iterations used in the hash value computation. It can be adjusted based on the desired level of security and the available computational resources. The function uses these inputs to compute a 24-byte (192-bit) hash. The final output will be a string as shown in Figure \ref{fig:bcrypt}.


    \subsection{Lattice-based Threshold Secret Sharing}
    \old{https://ieeexplore.ieee.org/stamp/stamp.jsp?tp=&arnumber=6994043}
    A secret sharing scheme is a technique used to distribute a secret among multiple participants by providing them with shares. A dealer distributes the shares through a secure channel to ensure that only authorized subsets of participants can recover the secret. A public key cryptosystem can represent the secure channel. Lattice-based cryptography is a promising area for designing new public key cryptosystems that can resist quantum computers \cite{10.1007/BFb0052231, 10.1007/BFb0054868, 10.1145/1568318.1568324}. Lattice-based Threshold Secret Sharing Scheme (LTSSS) is a variation of method used by Steinfeld et al. \cite{10.1007/978-3-540-30539-2_13}.
    
    The LTSSS consists of three phases: public parameter generation, share generation, and secret reconstruction.

    In the share generation phase, the dealer selects $n$ distinct $\mathbf{l}^{(i)}$ vectors of dimension $\mathit{m}$ uniformly at random, and an $\mathit{m}$-dimensional vector $\mathbf{a}$ is chosen such that its first component is assigned to the secret and the remaining $\mathit{(m - 1)}$ components are assigned random values. Then, the shares are computed by adding some noise $\epsilon_i$ to the inner product of $\mathbf{l}^{(i)}$ and $\mathbf{a}$, for each $i \in {1, 2, \ldots, n}$.

    In the secret reconstruction phase, a combiner (server) generates a $\mathit{(t + m)}$-dimensional lattice basis $\mathbf{M}$ by exploiting $t$ out of $n$ vectors $\mathbf{l}^{(i)}$, and a $\mathit{(t + m)}$-dimensional vector $\mathbf{t'}$ is generated using $t$ out of $n$ shares, which is close to a certain lattice point whose $\mathit{(t + 1)}^\text{th}$ component is a known fraction of the secret. By running an approximation algorithm to find the closest vector of the lattice generated by the basis $\mathbf{M}$ to $\mathbf{t'}$, the secret can be recovered.

    The LTSSS surpasses Shamir's approach as it embraces quantum security, avoids finite fields, supports diverse access structures, and offers efficient operations along with error-tolerance \cite{khorasgani2014lattice}. Information-theoretic security and flexibility make it an appealing alternative, despite being less mature in the field.
    
\section{Proposed Methodology}
\label{sec:proposed-methodology}

\begin{figure*}[!h]
    \centering\includegraphics[scale=0.7]{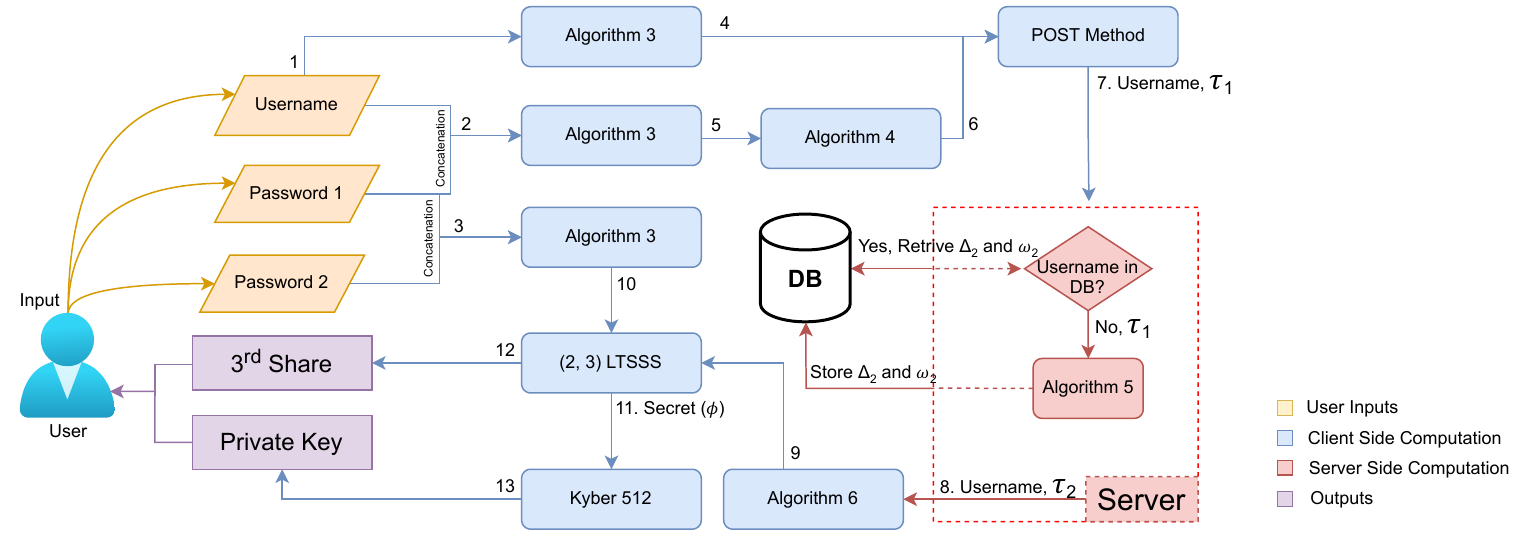}
    \caption{Working mechanism of the proposed solution}
    \label{fig:proposed-methodology}
\end{figure*}

The proposed methodology endeavors to heighten the security of digital asset wallets by amalgamating PQC with ZKP. As shown in Figure \ref{fig:proposed-methodology}, the proposed architecture comprises a client and a server, wherein the client engenders a secret key predicated on the user's username and passwords leveraging Lattice-based cryptography. The server, in turn, generates a random point on the lattice, which acts as a secret key and disseminates it with the client utilizing a LTSSS. In a secure manner, the client utilizes the generated secret key in tandem with ZKP to derive the private and public keys. Notably, this is achieved without any sensitive information being exposed to the server, thus bolstering the system's overall security.

\subsection{Algorithmic Insights}

\begin{algorithm}[!h]
\caption{Function to compute Hash value}\label{algo:hash_function}
\begin{algorithmic}
\Procedure{hash}{$message$}
\State $cost \gets 12$
\State $hashBcrypt \gets \Call{bcrypt}{cost, salt, message}$
\State $hashValue \gets \Call{SHA256}{hashBcrypt}$
\State \Return $hashValue$
\EndProcedure
\end{algorithmic}
\end{algorithm}

The procedure described in Algorithm \ref{algo:hash_function} is a function used to compute a hash value for a given message. It utilizes two cryptographic algorithms, bcrypt and SHA-256, to ensure the security and integrity of the resulting hash. The algorithm takes a message as input and begins by setting a cost parameter to 12. This cost parameter determines the computational effort required to compute the bcrypt hash. Higher values of cost increase the time and resources needed to compute the hash, thereby enhancing security against brute-force attacks. The bcrypt function applies a one-way hashing algorithm that incorporates the cost and salt to generate a bcrypt hash. The salt is a random value used to further strengthen the hash and protect against rainbow table attacks. The resulting bcrypt hash is then passed as input to the SHA-256 algorithm. It is designed to be computationally secure and resistant to collision attacks, where different inputs produce the same hash value.The computed SHA-256 hash value is assigned to the variable \quotes{hashValue} and returned as the output of the function.

\begin{algorithm}[h]
\caption{Client-side Lattice Computation} \label{algo:hash_to_lattice_point}
\begin{algorithmic}
\Procedure{LatticeComputation}{$\eta$}
    \State $d \gets 256$
    \State $q \gets 2^{14}$
    \State $s \gets 2d$
    \State $\sigma \gets \sqrt{\frac{d}{q}}$
    \State $binary\_string \gets \text{ConvertToBinary}(\eta, 16)$
    \State $\rho \gets []$
    \For{$i \gets 0$ \textbf{to} $d-1$ \textbf{step} $16$}
        \State $chunk \gets \text{binary\_string}[i:i+16]$
        \State $decimal\_value \gets \text{int}(chunk, 2)$
        \State $\rho.\text{append}(decimal\_value \mod q)$
    \EndFor
    \State $\Delta_1 \gets []$
    \For{$i \gets 0$ \textbf{to} $d-1$}
        \State $\Delta_1.\text{append}(\text{random.gaussian}(0, \sigma))$
    \EndFor
    \State $\omega_1 \gets []$
    \For{$i \gets 0$ \textbf{to} $s-1$}
        \State $\omega_1.\text{append}(\text{random.int}(0, 1))$
    \EndFor
    \State $\tau_1 \gets (\sum_{i=0}^{d-1} \rho[i] \cdot \omega_1[i]) + \left(\frac{q}{2} \cdot \sum_{i=0}^{d-1} \Delta_1[i]\right) \mod q$
    \State \textbf{return} $\tau_1, \Delta_1, \omega_1$
\EndProcedure
\end{algorithmic}
\end{algorithm}

The procedure presented in Algorithm \ref{algo:hash_to_lattice_point} performs client-side computations to transform a hexadecimal value into a lattice point within a lattice-based cryptography scheme. It accomplishes through the following steps:

\begin{itemize}
    \item Initialize the parameters: Set the dimension $d$ to 256, the modulus $q$ to $2^{14}$, and the lattice size $s$ to twice the dimension.
    \item Calculate the standard deviation $\sigma$: It is computed as the square root of the ratio between the dimension $d$ and the modulus $q$. This value is used in the following steps.
    \item Convert the hexadecimal value to a binary string: The input hexadecimal value is converted into a binary string representation with a fixed length of 16 bits.
    \item Prepare the $\rho$ vector: Iterate over the binary string in chunks of 16 bits and convert each chunk into a decimal value. Each decimal value is then reduced modulo $q$ and added to the $\rho$ vector, resulting in a vector of $d$ elements.
    \item Generate the $\Delta_1$ vector: Generate a vector $\Delta_1$ of $d$ elements by assigning random values from a Gaussian distribution with mean 0 and standard deviation $\sigma$.
    \item Generate the $\omega_1$ vector: Generate a binary vector $\omega_1$ of $s$ elements by randomly assigning either 0 or 1 to each element.
    \item Compute the lattice point $\tau_1$: Calculate $\tau_1$ as the sum of the element-wise multiplication between the $\rho$ and $\omega_1$ vectors, added to the product of half the modulus $q$ and the sum of the elements in the $\Delta_1$ vector. Finally, reduce the result modulo $q$.
    \item Return the lattice point: The algorithm returns the computed lattice point $\tau_1$, the vectors $\Delta_1$ and $\omega_1$ for future use.
\end{itemize}

\begin{algorithm}[h]
\caption{Server-side Lattice Computation} \label{algo:point_to_lattice_point}
\begin{algorithmic}
\Procedure{LatticeComputation}{$\tau_1$}
    \State $d \gets 256$
    \State $q \gets 2^{14}$
    \State $s \gets 2d$
    \State $\sigma \gets \sqrt{\frac{d}{q}}$
    \State $\Delta_2 \gets []$
    \For{$i \gets 0$ \textbf{to} $d-1$}
        \State $\Delta_2.\text{append}(\text{random.gaussian}(0, \sigma))$
    \EndFor
    \State $\omega_2 \gets []$
    \For{$i \gets 0$ \textbf{to} $s-1$}
        \State $\omega_2.\text{append}(\text{random.int}(0, 1))$
    \EndFor
    \State $\tau_2 \gets (\sum_{i=0}^{d-1} \tau_1[i] \cdot \omega_2[i]) + \left(\frac{q}{2} \cdot \sum_{i=0}^{d-1} \Delta_2[i]\right) \mod q$
    \State \textbf{return} $\tau_2, \Delta_2, \omega_2$
\EndProcedure
\end{algorithmic}
\end{algorithm}

The procedure in Algorithm \ref{algo:point_to_lattice_point} is similar to Algorithm \ref{algo:hash_to_lattice_point}, performs server-side computations to transform a given lattice point represented by the $\tau_1$ vector into a new lattice point. The algorithm initializes the parameters, including the dimension $d$, modulus $q$, and lattice size $s$, which define the properties of the lattice. It then generates two vectors, $\Delta_2$ and $\omega_2$, of size $d$ and $s$ respectively. The $\Delta_2$ vector is created by assigning random values from a Gaussian distribution with mean 0 and standard deviation $\sigma$, which is computed as the square root of $d/q$. The $\omega_2$ vector is formed by randomly selecting either 0 or 1 for each element. It proceeds to compute the new lattice point $\tau_2$. This is accomplished by performing element-wise multiplication between the $\tau_1$ vector and the $\omega_2$ vector, followed by summing the products. Additionally, half the modulus $q$ is multiplied by the sum of the elements in the $\Delta_2$ vector. The resulting values are combined and reduced modulo $q$ to obtain the new lattice point $\tau_2$. Finally, the algorithm returns the computed new lattice point $\tau_2$, along with the $\Delta_2$ and $\omega_2$ vectors. These values are used in subsequent steps of the proposed system.

\begin{algorithm}[h]
\caption{Real Lattice Point Generation}\label{alg:original_lattice_point}
\begin{algorithmic}
\Procedure{RealLatticePoint}{$\tau_2, \Delta_1, \omega_1$}
    \State $d \gets 256$
    \State $q \gets 2^{14}$
    \State $dot\_product \gets 0$
    \For{$i \gets 0$ to $\text{length}(\omega_1)-1$}
        \State $dot\_product \gets dot\_product + \omega_1[i] \times \Delta_1[i]$
    \EndFor
    \State $q_0 \gets \lfloor q/2 \rfloor$
    \State $diff \gets (\tau_2 - q_0 \times dot\_product) \mod q$
    \State $inv\_q_0 \gets \text{ExtendedEuclidean}(q_0, q)$
    \State $result \gets (inv\_q_0 \times diff) \mod q$
    \State $\rho \gets []$
    \For{$i \gets 0$ to $d-1$ step $16$}
        \State $chunk \gets \text{Substring}(result, i, i+15)$
        \State $decimal\_value \gets \text{ConvertToDecimal}(chunk, 2)$
        \State $\rho.\text{Append}(decimal\_value)$
    \EndFor
    \State \textbf{return} $\rho$
\EndProcedure
\end{algorithmic}
\end{algorithm}

The procedure presented in Algorithm \ref{alg:original_lattice_point} is used to generate the original lattice point $\rho$ from the provided parameters $\tau_2$, $\Delta_1$, and $\omega_1$. The algorithm performs the following steps:

\begin{itemize}
    \item Initialize the dimension $d$ to $256$ and modulus $q$ to $2^{14}$, which defines the properties of the lattice.
    \item Compute the dot product between the vectors $\Delta_1$ and $\omega_1$ by iterating over their elements and accumulating the product of corresponding elements.
    \item Compute $q_0$ as half of the modulus $q$.
    \item Calculate the difference between $\tau_2$ and $q_0$ times the dot product. Reduce the result modulo $q$ to ensure it falls within the lattice.
    \item Compute the modular inverse $inv\_q_0$ of $q_0$ using the Extended Euclidean algorithm.
    \item Multiply $inv\_q_0$ with the difference calculated earlier and reduce the result modulo $q$. This step ensures that the final result falls within the lattice and can be represented as a valid lattice point.
    \item Construct the $\rho$ vector by splitting the final result into chunks of 16 bits.
    \item Convert each chunk from binary to decimal representation.
    \item Append the decimal values to the $\rho$ vector.
    \item Return the $\rho$ vector, representing the original lattice point.
\end{itemize}


\subsection{Working Mechanism}
The proposed methodology is divided into three distinct components, namely the client-side computation, server-side computation and client-side key generation.

\textbf{Client-side Computation:}
\begin{enumerate}
    \item Compute $\eta$ by applying the hash function (see Algorithm \ref{algo:hash_function}) to the username and password1.
    \item Compute $\mu$ by applying the hash function (see Algorithm \ref{algo:hash_function}) to password1 and password2.
    \item Compute $\tau_1$, $\Delta_1$, and $\omega_1$ by using the hash-to-lattice-point algorithm (see Algorithm \ref{algo:hash_to_lattice_point}) with $\eta$ as input.
    \item Send a POST request to the server, including the hashed username and $\tau_1$.
\end{enumerate}

\textbf{Server-side Computation:}
\begin{enumerate}
    \item If the hashed username exists in the server's database, retrieve the stored values $\Delta_2$ and $\omega_2$ associated with it.
    \item If the hashed username does not exist, compute $\tau_2$ by using the point-to-lattice-point algorithm (see Algorithm \ref{algo:point_to_lattice_point}) with $\tau_1$ as input.
    \item Store the computed $\Delta_2$ and $\omega_2$ in the database, mapping them to the hashed username.
    \item Send a response to the client, including $\tau_2$.
\end{enumerate}

\textbf{Client-side Key Generation:}
\begin{enumerate}
    \item Compute $\rho$ by using the real lattice point algorithm (see Algorithm \ref{alg:original_lattice_point}) with $\tau_2$, $\Delta_1$, and $\omega_1$ as input.
    \item Generate a secret $\phi$ by using $\rho$ and $\mu$ as two shares of a (2, 3) LTSSS.
    \item Generate a third share from the secret for backup purposes.
    \item Generate a private and public key pair using the Kyber512 algorithm, with $\phi$ as input.
\end{enumerate}

On the client-side, the computation begins with the application of Algorithm \ref{algo:hash_function} i.e., a hash function to the username and password1, resulting in a hashed value called $\eta$. Additionally, another hash computation is performed using password1 and password2, producing a hashed value called $\mu$. These hash functions ensure the confidentiality and integrity of sensitive information. Furthermore, the client applies Algorithm \ref{algo:hash_to_lattice_point} i.e., a lattice computation function to $\eta$, resulting in the computation of $\tau_1$, $\Delta_1$, and $\omega_1$. These values represent points on a lattice structure and serve as the foundation for subsequent operations. The client then sends a POST request to the server, transmitting the hashed username and $\tau_1$. This request establishes a communication channel between the client and server.

On the server-side, the server checks if the hashed username exists in its database. If it does, the server retrieves the corresponding stored values $\Delta_2$ and $\omega_2$. These values ensure consistency in the subsequent computations. If the hashed username does not exist, the server computes $\tau_2$ using Algorithm \ref{algo:point_to_lattice_point} i.e., the server-side lattice computation with $\tau_1$ as input. The server stores the computed $\Delta_2$ and $\omega_2$ in its database, associating them with the hashed username for future reference. A response is then sent back to the client, including $\tau_2$, establishing a secure channel for further communication.

At the end, on the client-side, the client computes $\rho$ by using the Algorithm \ref{alg:original_lattice_point} i.e., real lattice point generation function with $\tau_2$, $\Delta_1$, and $\omega_1$ as input. This computation converts the received lattice point into a usable form. The client generates a backup key or third share, a secret $\phi$ by utilizing $\rho$ and $\mu$ as shares of a (2, 3) LTSSS. This ensures that multiple shares are needed to reconstruct the secret, enhancing security. A third share is generated for backup purposes. Additionally, the client generates a private and public key pair using the Kyber512 algorithm with $\phi$ as input. This key pair forms the foundation for secure data exchange.

Users can securely store the backup key using methods such as offline storage (e.g., printed paper), secure local storage, or cloud storage. This backup key ensures access to the private key and the ability to perform essential transactions even in scenarios where the server experiences downtime or disruptions.

The proposed method also incorporates a noteworthy feature that allows users to rekey their private key while retaining the same username and passwords. When users opt for rekeying, the server assigns new random values for $\Delta_2$ and $\omega_2$, which are subsequently utilized in generating a fresh private key. It is essential to emphasize that the values $\Delta_2$ and $\omega_2$ stored in the database hold no valuable information pertaining to the private key itself; rather, they serve as inputs for the key generation process. The inclusion of this rekeying functionality offers an elevated level of security and flexibility to users. In the event of suspected private key compromise or a desire to enhance security measures, users can effortlessly initiate the rekeying process. By doing so, any potential vulnerabilities associated with the previous private key are effectively mitigated. The freshly generated private key, derived from the updated $\Delta_2$ and $\omega_2$ values, operates independently from its predecessor and introduces an enhanced level of security. By separating the private key generation process from the username and passwords, the proposed method ensures that even if the stored $\Delta_2$ and $\omega_2$ values are compromised, they hold no information that could be exploited to reveal the private key. This reinforces the safeguarding of user assets and bolsters the overall security of the wallet.

\section{Results}
\label{sec:results}

The proposed method generates several key results that contribute to its superiority over existing wallets and enhance its security against quantum computers.

\begin{itemize}
    \item \textbf{User-Friendly Authentication:} Instead of relying on complex recovery phrases, this method uses familiar credentials, namely a username and password, for authentication. This approach simplifies the user experience and makes it easier for individuals to access their wallets securely.
    \item \textbf{Quantum-Resistance:} The method addresses the emerging threat of quantum computers by employing quantum-resistant cryptographic algorithms. The wallet generates private and public keys using the Kyber512 algorithm, which is designed to withstand attacks from quantum computers. This ensures that the generated keys remain secure even in the face of advancements in quantum computing technology.
    \item \textbf{Robust Key Size:} The wallet generates a Kyber512 private key, which has a key size of 512 bits. The larger key size provides significantly higher cryptographic strength compared to the 128 or 256-bit keys commonly used in recovery phrase-based wallets. This increased key size contributes to the wallet's overall security and makes it more resistant to brute-force and cryptographic attacks.
    \item \textbf{Rekeying Capability:} One notable advantage of this method is the ability to rekey the private key using the same username and password. In traditional recovery phrase-based wallets, the recovery phrase is the sole means of regaining access to the wallet. However, in this method, users can rekey their private key by providing their username and password. This feature offers convenience and flexibility, allowing users to regain access to their wallets using familiar credentials without solely relying on a recovery phrase.
\end{itemize}

\section{Implications}
\label{sec:implications}
The utilization of LWE-based cryptography within the cryptocurrency wallet algorithm presents a compelling avenue for achieving a high level of security against known attacks targeting the LWE problem. The LWE problem has exhibited formidable resistance to a diverse range of attacks, spanning both classical and quantum algorithms, rendering it an attractive option for cryptographic applications. The incorporation of Secure Sockets Layer (SSL) and Multi-Factor Authentication (MFA) serves as an effective deterrent against the vulnerability posed by Man-in-The-Middle (MiTM) attacks. SSL serves to facilitate a secure and encrypted communication channel between two entities, thereby impeding malevolent actors from intercepting and manipulating data while in transit. MFA, on the other hand, amplifies the security of user credentials by fortifying the authentication process through the stipulation of at least two identification modes prior to access being granted to their account.

Nevertheless, It is crucial to acknowledge that although these measures are efficacious, they are not entirely impervious to the risks of MITM attacks. The possibility of exploiting SSL or MFA implementation vulnerabilities, or employing tactics such as phishing by perpetrators, remains a potential concern. The comprehensive security of the wallet algorithm is also reliant on the security of other constituent elements, including the authentication mechanism and the storage of shared secrets. Notably, the deployment of a two-password system for generating the shared secret may engender certain usability challenges that necessitate careful consideration.

Users are required to remember two distinct passwords for the two stages of password hashing, presenting a plausible usability barrier. This may result in users struggling to remember complex passwords for both stages or forgetting one of the passwords entirely, ultimately resulting in self-lockout from the wallet. It is crucial to design password requirements and recovery mechanisms that balance security and usability to mitigate this issue. Despite these challenges, the two-password system implemented in the wallet algorithm may be more user-friendly than current wallet systems that rely on 12/24 word seed-phrases. Such phrases may be difficult for users to recall and susceptible to brute-force attacks if not chosen carefully.

Apart from usability challenges, it is also important to consider the potential for attacks on the LWE-based cryptography in the future. While the LWE problem is currently deemed to be secure against a wide range of attacks, the possibility of new attacks or quantum computing advances rendering the system vulnerable cannot be entirely precluded. To address this issue, the wallet algorithm incorporates additional security measures, such as PQC and ZKP. PQC techniques have been devised to withstand attacks by quantum computers, which can break many traditional cryptographic schemes. Meanwhile, ZKP techniques enable cryptographic proofs to be furnished without revealing sensitive information, ensuring the continued security of the system, even if the attacker attains partial access to the system.

The use of LWE-based cryptography in the cryptocurrency wallet algorithm represents a promising approach for achieving high security. Nevertheless, the system's overall security is contingent upon the security of multiple components, including the password authentication mechanism and the storage of shared secrets. It is imperative to balance usability and security when designing these elements to ensure that the system is secure and user-friendly. Furthermore, it is vital to continually monitor and update the system's security in response to emerging threats and advances in cryptographic research.

\section{Conclusion}
\label{sec:conclusion}
Our proposed algorithm for improving the security of cryptocurrency wallets using LWE-based cryptography offers several advantages over existing solutions. Unlike many current wallets that rely on a 12/24 word seed-phrase, our algorithm uses a two-password system for generating the shared secret. This not only provides stronger security against quantum attacks, but also makes the wallet more user-friendly by reducing the risk of users losing their recovery phrase. Our algorithm also offers the added advantage of rekeying the private key using the same username and password, further enhancing the wallet's security.

Moreover, using LWE-based cryptography provides a high level of security against known attacks. The algorithm also provides post-quantum security using PQC and ZKP, ensuring the wallet remains secure even in future attacks on LWE-based cryptography. However, we recognize that using a two-password system may be inconvenient for some users, as they need to remember two passwords and ensure they are sufficiently complex and secure. To address this, we plan to explore alternative approaches that improve the user experience while maintaining a high level of security, such as using a single password or biometric based authentication, social recovery mechanisms.


\bibliographystyle{IEEEtran}
\bibliography{bibg}
\end{document}

%% file: Tables/PQC-Comparision.tex
\begin{table}[ht]
	\centering
	\caption{Comparison of Key Sizes, Ciphertext, and Signature Sizes for Various Post-Quantum Cryptography Algorithms at a 128-bit Security Level}
	\label{tab:PQC-Comparision}
        \renewcommand{\arraystretch}{1.5}
	\resizebox{\linewidth}{!}{
		\begin{tabular}{l|llll}
			\hline
			Algorithm                               & Type           & Public Key & Private Key & signature \\ \hline
			NTRU Encrypt                       & Lattice        & 766.25 B   & 842.875 B   & -         \\
			Streamlined NTRU Prime                  & Lattice        & 154 B      & -           & -         \\
			BLISS-II                                & Lattice        & 7 KB       & 2 KB        & 5 KB      \\
			Rainbow                            & Multivariate   & 124 KB     & 95 KB       & -         \\
			SPHINCS                            & Hash Signature & 1 KB       & 1 KB        & 41 KB     \\
			SPHINCS+                            & Hash Signature & 32 B       & 64 B        & 8 KB      \\
			GLP-Variant GLYPH Signature     & Ring-LWE       & 2 KB       & 0.4 KB      & 1.8 KB    \\
			NewHope                             & Ring-LWE       & 2 KB       & 2 KB        & -         \\
			Goppa-based McEliece                & Code-based     & 1 MB       & 11.5 KB     & -         \\
			Quasi-cyclic MDPC-based McEliece    & Code-based     & 1,232 B    & 2,464 B     & -         \\
			Random Linear Code based encryption & RLCE           & 115 KB     & 3 KB        & -         \\
			SIDH                                & Isogeny        & 564 B      & 48 B        & -         \\
			SIDH (compressed keys)              & Isogeny        & 330 B      & 48 B        & -         \\
			3072-bit Discrete Log                   & not PQC        & 384 B      & 32 B        & 96 B      \\
			256-bit Elliptic Curve                  & not PQC        & 32 B       & 32 B        & 65 B      \\ \hline
		\end{tabular}
	}
\end{table}

%% file: main.bbl
\begin{thebibliography}{10}
\providecommand{\url}[1]{#1}
\csname url@samestyle\endcsname
\providecommand{\newblock}{\relax}
\providecommand{\bibinfo}[2]{#2}
\providecommand{\BIBentrySTDinterwordspacing}{\spaceskip=0pt\relax}
\providecommand{\BIBentryALTinterwordstretchfactor}{4}
\providecommand{\BIBentryALTinterwordspacing}{\spaceskip=\fontdimen2\font plus
\BIBentryALTinterwordstretchfactor\fontdimen3\font minus
  \fontdimen4\font\relax}
\providecommand{\BIBforeignlanguage}[2]{{%
\expandafter\ifx\csname l@#1\endcsname\relax
\typeout{** WARNING: IEEEtran.bst: No hyphenation pattern has been}%
\typeout{** loaded for the language `#1'. Using the pattern for}%
\typeout{** the default language instead.}%
\else
\language=\csname l@#1\endcsname
\fi
#2}}
\providecommand{\BIBdecl}{\relax}
\BIBdecl

\bibitem{satoshi2009bitcoin}
S.~Nakamoto, ``Bitcoin: A peer-to-peer electronic cash system,''
  \emph{Cryptography Mailing list at https://metzdowd.com}, 03 2009.

\bibitem{erinle2023sok}
Y.~Erinle, Y.~Kethepalli, Y.~Feng, and J.~Xu, ``Sok: Design, vulnerabilities
  and defense of cryptocurrency wallets,'' \emph{arXiv preprint
  arXiv:2307.12874}, 2023.

\bibitem{shor1997polynomial}
\BIBentryALTinterwordspacing
P.~W. Shor, ``Polynomial-time algorithms for prime factorization and discrete
  logarithms on a quantum computer,'' \emph{SIAM Journal on Computing},
  vol.~26, no.~5, pp. 1484--1509, 1997. [Online]. Available:
  \url{https://doi.org/10.1137/S0097539795293172}
\BIBentrySTDinterwordspacing

\bibitem{bernstein2009quantum}
\BIBentryALTinterwordspacing
D.~J. Bernstein, \emph{Introduction to post-quantum cryptography}.\hskip 1em
  plus 0.5em minus 0.4em\relax Berlin, Heidelberg: Springer Berlin Heidelberg,
  2009, pp. 1--14. [Online]. Available:
  \url{https://doi.org/10.1007/978-3-540-88702-7_1}
\BIBentrySTDinterwordspacing

\bibitem{mangipudi2022uncovering}
E.~V. Mangipudi, U.~Desai, M.~Minaei, M.~Mondal, and A.~Kate, ``Uncovering
  impact of mental models towards adoption of multi-device crypto-wallets,''
  \emph{Cryptology ePrint Archive}, 2022.

\bibitem{he2018social}
S.~He, Q.~Wu, X.~Luo, Z.~Liang, D.~Li, H.~Feng, H.~Zheng, and Y.~Li, ``A
  social-network-based cryptocurrency wallet-management scheme,'' \emph{IEEE
  Access}, vol.~6, pp. 7654--7663, 2018.

\bibitem{pal2021key}
O.~Pal, B.~Alam, V.~Thakur, and S.~Singh, ``Key management for blockchain
  technology,'' \emph{ICT express}, vol.~7, no.~1, pp. 76--80, 2021.

\bibitem{he2019novel}
X.~He, J.~Lin, K.~Li, and X.~Chen, ``A novel cryptocurrency wallet management
  scheme based on decentralized multi-constrained derangement,'' \emph{IEEE
  Access}, vol.~7, pp. 185\,250--185\,263, 2019.

\bibitem{courtois2017stealth}
N.~T. Courtois and R.~Mercer, ``Stealth address and key management techniques
  in blockchain systems,'' in \emph{ICISSP 2017-Proceedings of the 3rd
  International Conference on Information Systems Security and Privacy}, 2017,
  pp. 559--566.

\bibitem{Belchior2022}
R.~Belchior, A.~Vasconcelos, S.~Guerreiro, and M.~Correia, ``{A Survey on
  Blockchain Interoperability: Past, Present, and Future Trends},'' \emph{ACM
  Computing Surveys}, vol.~54, no.~8, 2022.

\bibitem{Chen2020b}
H.~Chen, M.~Pendleton, L.~Njilla, and S.~Xu, ``{A Survey on Ethereum Systems
  Security: Vulnerabilities, Attacks, and Defenses},'' \emph{ACM Computing
  Surveys}, vol.~53, no.~3, 2020.

\bibitem{suratkar2020cryptocurrency}
S.~Suratkar, M.~Shirole, and S.~Bhirud, ``Cryptocurrency wallet: A review,'' in
  \emph{2020 4th International Conference on Computer, Communication and Signal
  Processing (ICCCSP)}.\hskip 1em plus 0.5em minus 0.4em\relax IEEE, 2020, pp.
  1--7.

\bibitem{eskandari2018first}
S.~Eskandari, J.~Clark, D.~Barrera, and E.~Stobert, ``A first look at the
  usability of bitcoin key management,'' \emph{arXiv preprint
  arXiv:1802.04351}, 2018.

\bibitem{gotte2021tech}
J.~S. G{\"o}tte and B.~Scheuermann, ``Tech report: Inerial hsms thwart advanced
  physical attacks,'' \emph{Cryptology ePrint Archive}, 2021.

\bibitem{shbair2021hsm}
W.~M. Shbair, E.~Gavrilov, and R.~State, ``Hsm-based key management solution
  for ethereum blockchain,'' in \emph{2021 IEEE International Conference on
  Blockchain and Cryptocurrency (ICBC)}.\hskip 1em plus 0.5em minus 0.4em\relax
  IEEE, 2021, pp. 1--3.

\bibitem{khadzhi2020method}
A.~S. Khadzhi, S.~V. Zareshin, and O.~V. Tarakanov, ``A method for analyzing
  the activity of cold wallets and identifying abandoned cryptocurrency
  wallets,'' in \emph{2020 IEEE Conference of Russian Young Researchers in
  Electrical and Electronic Engineering (EIConRus)}.\hskip 1em plus 0.5em minus
  0.4em\relax IEEE, 2020, pp. 1974--1977.

\bibitem{khan2019security}
A.~G. Khan, A.~H. Zahid, M.~Hussain, and U.~Riaz, ``Security of cryptocurrency
  using hardware wallet and qr code,'' in \emph{2019 International Conference
  on Innovative Computing (ICIC)}.\hskip 1em plus 0.5em minus 0.4em\relax IEEE,
  2019, pp. 1--10.

\bibitem{eyal2022cryptocurrency}
I.~Eyal, ``On cryptocurrency wallet design,'' in \emph{3rd International
  Conference on Blockchain Economics, Security and Protocols (Tokenomics
  2021)}.\hskip 1em plus 0.5em minus 0.4em\relax Schloss
  Dagstuhl-Leibniz-Zentrum f{\"u}r Informatik, 2022.

\bibitem{di2020characteristics}
M.~Di~Angelo and G.~Salzer, ``Characteristics of wallet contracts on
  ethereum,'' in \emph{2020 2nd Conference on Blockchain Research \&
  Applications for Innovative Networks and Services (BRAINS)}.\hskip 1em plus
  0.5em minus 0.4em\relax IEEE, 2020, pp. 232--239.

\bibitem{homoliak2018air}
I.~Homoliak, D.~Breitenbacher, A.~Binder, and P.~Szalachowski, ``An air-gapped
  2-factor authentication for smart-contract wallets,'' \emph{arXiv preprint
  arXiv:1812.03598}, 2018.

\bibitem{rezaeighaleh2019deterministic}
H.~Rezaeighaleh and C.~C. Zou, ``Deterministic sub-wallet for
  cryptocurrencies,'' in \emph{2019 IEEE International Conference on Blockchain
  (Blockchain)}.\hskip 1em plus 0.5em minus 0.4em\relax IEEE, 2019, pp.
  419--424.

\bibitem{chen2020survey}
H.~Chen, M.~Pendleton, L.~Njilla, and S.~Xu, ``A survey on ethereum systems
  security: Vulnerabilities, attacks, and defenses,'' \emph{ACM Computing
  Surveys (CSUR)}, vol.~53, no.~3, pp. 1--43, 2020.

\bibitem{li2020survey}
X.~Li, P.~Jiang, T.~Chen, X.~Luo, and Q.~Wen, ``A survey on the security of
  blockchain systems,'' \emph{Future Generation Computer Systems}, vol. 107,
  pp. 841--853, 2020.

\bibitem{guo2022survey}
H.~Guo and X.~Yu, ``A survey on blockchain technology and its security,''
  \emph{Blockchain: Research and Applications}, vol.~3, no.~2, p. 100067, 2022.

\bibitem{zamani2020security}
E.~Zamani, Y.~He, and M.~Phillips, ``On the security risks of the blockchain,''
  \emph{Journal of Computer Information Systems}, vol.~60, no.~6, pp. 495--506,
  2020.

\bibitem{Urien2021}
P.~Urien, ``{Innovative Countermeasures to Defeat Cyber Attacks Against
  Blockchain Wallets},'' \emph{2021 5th Cyber Security in Networking
  Conference, CSNet 2021}, pp. 49--54, 2021.

\bibitem{bui2019pitfalls}
T.~Bui, S.~P. Rao, M.~Antikainen, and T.~Aura, ``Pitfalls of open architecture:
  How friends can exploit your cryptocurrency wallet,'' in \emph{Proceedings of
  the 12th European Workshop on Systems Security}, 2019, pp. 1--6.

\bibitem{andryukhin2019phishing}
A.~Andryukhin, ``Phishing attacks and preventions in blockchain based
  projects,'' in \emph{2019 International Conference on Engineering
  Technologies and Computer Science (EnT)}.\hskip 1em plus 0.5em minus
  0.4em\relax IEEE, 2019, pp. 15--19.

\bibitem{gershon2013qubit}
\BIBentryALTinterwordspacing
E.~Gershon, ``New qubit control bodes well for future of quantum computing,'' 1
  2013. [Online]. Available:
  \url{https://phys.org/news/2013-01-qubit-bodes-future-quantum.html}
\BIBentrySTDinterwordspacing

\bibitem{spectrum2008quantum}
\BIBentryALTinterwordspacing
I.~Spectrum, ``How quantum computers threaten our current cryptography system
  and what we can do about it,'' 11 2008. [Online]. Available:
  \url{https://spectrum.ieee.org/qa-with-postquantum-computing-cryptography-researcher-jintai-ding}
\BIBentrySTDinterwordspacing

\bibitem{bernstein2009sharcs}
\BIBentryALTinterwordspacing
D.~J. Bernstein, ``Cost analysis of hash collisions: Will quantum computers
  make sharcs obsolete?'' 5 2009. [Online]. Available:
  \url{https://cr.yp.to/hash/collisioncost-20090517.pdf}
\BIBentrySTDinterwordspacing

\bibitem{bernstein2010gvm}
\BIBentryALTinterwordspacing
------, ``Grover vs. mceliece,'' 3 2010. [Online]. Available:
  \url{https://cr.yp.to/codes/grovercode-20100303.pdf}
\BIBentrySTDinterwordspacing

\bibitem{cryptoeprint:2014/070}
\BIBentryALTinterwordspacing
C.~Peikert, ``Lattice cryptography for the internet,'' Cryptology ePrint
  Archive, Paper 2014/070, 2014, \url{https://eprint.iacr.org/2014/070}.
  [Online]. Available: \url{https://eprint.iacr.org/2014/070}
\BIBentrySTDinterwordspacing

\bibitem{10.1007/978-3-642-33027-8_31}
T.~G{\"u}neysu, V.~Lyubashevsky, and T.~P{\"o}ppelmann, ``Practical
  lattice-based cryptography: A signature scheme for embedded systems,'' in
  \emph{Cryptographic Hardware and Embedded Systems -- CHES 2012}, E.~Prouff
  and P.~Schaumont, Eds.\hskip 1em plus 0.5em minus 0.4em\relax Berlin,
  Heidelberg: Springer Berlin Heidelberg, 2012, pp. 530--547.

\bibitem{10.1007/978-3-662-46803-6_24}
J.~Zhang, Z.~Zhang, J.~Ding, M.~Snook, and {\"O}.~Dagdelen, ``Authenticated key
  exchange from ideal lattices,'' in \emph{Advances in Cryptology - EUROCRYPT
  2015}, E.~Oswald and M.~Fischlin, Eds.\hskip 1em plus 0.5em minus 0.4em\relax
  Berlin, Heidelberg: Springer Berlin Heidelberg, 2015, pp. 719--751.

\bibitem{cryptoeprint:2013/383}
\BIBentryALTinterwordspacing
L.~Ducas, A.~Durmus, T.~Lepoint, and V.~Lyubashevsky, ``Lattice signatures and
  bimodal gaussians,'' Cryptology ePrint Archive, Paper 2013/383, 2013,
  \url{https://eprint.iacr.org/2013/383}. [Online]. Available:
  \url{https://eprint.iacr.org/2013/383}
\BIBentrySTDinterwordspacing

\bibitem{10.1145/2535925}
\BIBentryALTinterwordspacing
V.~Lyubashevsky, C.~Peikert, and O.~Regev, ``On ideal lattices and learning
  with errors over rings,'' \emph{J. ACM}, vol.~60, no.~6, nov 2013. [Online].
  Available: \url{https://doi.org/10.1145/2535925}
\BIBentrySTDinterwordspacing

\bibitem{Augot2015InitialRO}
D.~Augot, L.~Batina, D.~J. Bernstein, J.~W. Bos, J.~A. Buchmann, W.~Castryck,
  O.~Dunkelman, T.~G{\"u}neysu, S.~Gueron, A.~H{\"u}lsing, T.~Lange,
  C.~Rechberger, P.~Schwabe, N.~Sendrier, F.~Vercauteren, and B.-Y. Yang,
  ``Initial recommendations of long-term secure post-quantum systems,'' 2015.

\bibitem{lixie2017zeroknowledge}
\BIBentryALTinterwordspacing
Lixie, ``Zero-knowledge proofs explained: Part 1,'' 10 2021. [Online].
  Available:
  \url{https://www.expressvpn.com/blog/zero-knowledge-proofs-explained/}
\BIBentrySTDinterwordspacing

\bibitem{10.1145/62212.62222}
\BIBentryALTinterwordspacing
M.~Blum, P.~Feldman, and S.~Micali, ``Non-interactive zero-knowledge and its
  applications,'' in \emph{Proceedings of the Twentieth Annual ACM Symposium on
  Theory of Computing}, ser. STOC '88.\hskip 1em plus 0.5em minus 0.4em\relax
  New York, NY, USA: Association for Computing Machinery, 1988, p. 103–112.
  [Online]. Available: \url{https://doi.org/10.1145/62212.62222}
\BIBentrySTDinterwordspacing

\bibitem{Wu2014}
\BIBentryALTinterwordspacing
H.~Wu and F.~Wang, ``A survey of noninteractive zero knowledge proof system and
  its applications,'' \emph{The Scientific World Journal}, vol. 2014, p.
  560484, May 2014. [Online]. Available:
  \url{https://doi.org/10.1155/2014/560484}
\BIBentrySTDinterwordspacing

\bibitem{10.1007/0-387-34805-0_60}
J.-J. Quisquater, M.~Quisquater, M.~Quisquater, M.~Quisquater, L.~Guillou,
  M.~A. Guillou, G.~Guillou, A.~Guillou, G.~Guillou, and S.~Guillou, ``How to
  explain zero-knowledge protocols to your children,'' in \emph{Advances in
  Cryptology --- CRYPTO' 89 Proceedings}, G.~Brassard, Ed.\hskip 1em plus 0.5em
  minus 0.4em\relax New York, NY: Springer New York, 1990, pp. 628--631.

\bibitem{10.1007/3-540-39118-5_13}
D.~Chaum, J.-H. Evertse, and J.~van~de Graaf, ``An improved protocol for
  demonstrating possession of discrete logarithms and some generalizations,''
  in \emph{Advances in Cryptology --- EUROCRYPT' 87}, D.~Chaum and W.~L. Price,
  Eds.\hskip 1em plus 0.5em minus 0.4em\relax Berlin, Heidelberg: Springer
  Berlin Heidelberg, 1988, pp. 127--141.

\bibitem{10.1007/BFb0052231}
O.~Goldreich, S.~Goldwasser, and S.~Halevi, ``Public-key cryptosystems from
  lattice reduction problems,'' in \emph{Advances in Cryptology --- CRYPTO
  '97}, B.~S. Kaliski, Ed.\hskip 1em plus 0.5em minus 0.4em\relax Berlin,
  Heidelberg: Springer Berlin Heidelberg, 1997, pp. 112--131.

\bibitem{10.1007/BFb0054868}
J.~Hoffstein, J.~Pipher, and J.~H. Silverman, ``Ntru: A ring-based public key
  cryptosystem,'' in \emph{Algorithmic Number Theory}, J.~P. Buhler, Ed.\hskip
  1em plus 0.5em minus 0.4em\relax Berlin, Heidelberg: Springer Berlin
  Heidelberg, 1998, pp. 267--288.

\bibitem{10.1145/1568318.1568324}
\BIBentryALTinterwordspacing
O.~Regev, ``On lattices, learning with errors, random linear codes, and
  cryptography,'' \emph{J. ACM}, vol.~56, no.~6, sep 2009. [Online]. Available:
  \url{https://doi.org/10.1145/1568318.1568324}
\BIBentrySTDinterwordspacing

\bibitem{10.1007/978-3-540-30539-2_13}
R.~Steinfeld, H.~Wang, and J.~Pieprzyk, ``Lattice-based threshold-changeability
  for standard shamir secret-sharing schemes,'' in \emph{Advances in Cryptology
  - ASIACRYPT 2004}, P.~J. Lee, Ed.\hskip 1em plus 0.5em minus 0.4em\relax
  Berlin, Heidelberg: Springer Berlin Heidelberg, 2004, pp. 170--186.

\bibitem{khorasgani2014lattice}
H.~A. Khorasgani, S.~Asaad, T.~Eghlidos, and M.~Aref, ``A lattice-based
  threshold secret sharing scheme,'' in \emph{2014 11th International ISC
  Conference on Information Security and Cryptology}.\hskip 1em plus 0.5em
  minus 0.4em\relax IEEE, 2014, pp. 173--179.

\end{thebibliography}
